\documentclass{PoS}
\usepackage{subfigure}
\usepackage{mathtools}
\usepackage{wrapfig}
\title{Pion form factor from 2+1 dynamical flavor lattice QCD using the O(a) improved Wilson-clover quark formalism}

\ShortTitle{The pion form factor from 2+1 dynamical flavors lattice QCD}

\author{\speaker{Oanh Hoang Nguyen} for PACS-CS Collaboration\\
        Center for Computational Sciences and \\
	Graduate School of Pure and Applied Sciences\\
	University of Tsukuba\\
	Tsukuba, Ibaraki 305-3577\\
	Japan\\
        E-mail: \email{nhoanh@ccs.tsukuba.ac.jp}}

\abstract{We present a status report of our calculation of the electromagnetic form factor of the pion on the PACS-CS gauge field configurations 
generated using Iwasaki gauge action and Wilson-clover quark action for 2 + 1 flavors of light dynamical quarks.  
The technique of twiseted boundary conditon is employed to explore the form factor for small momentum transfer. 
The $q^2$ behavior of the form factor and its light quark mass dependence is examined.}

\FullConference{The XXVII International Symposium on Lattice Field Theory - LAT2009\\
		 July 26-31 2009\\
		 Peking University, Beijing, China}

\begin{document}

\section{Introduction}
The electromagnetic form factor of the pion $G_\pi(Q^2)$ 
is a simple and yet interesting quantity to explore hadronic structure using lattice QCD.  At small momentum transfer, 
{\it e.g.}, at $Q^2<0.3$~GeV$^2$, the form factor is well parametrized in terms of the mean-square charge radius whose experimental value 
is $\left<r^2\right>_{exp} = 0.452(11)$ fm$^2$.
Since this is a well measured quantity, lattice calculation of $\left<r^2\right>$ is a good check for the validity of lattice QCD.

In the framework of ChPT up to NLO, $\left<r^2\right>$ was calculated by Gasser and Leutwyler in their famous work~\cite{gasserleutwyler}:
\begin{equation}
\left< r^2\right>_{SU(2),NLO} = -\frac{1}{f^2} \left( 12l^r_6 + \frac{1}{8\pi^2} + \frac{1}{8\pi^2} log\frac{m^2_\pi}{\mu^2} \right).
\end{equation}
There is no $m^2_\pi$ suppression factor in front of the logarithm in this formula, 
and thus it predicts a rapid growth of the mean-square charge radius at small pion mass.
There have been a number of efforts in the lattice QCD community to verify this behavior including recent studies with dynamical 
quarks \cite{brommel,RBC,TMC,kaneko} with the pion mass as light as about 300 MeV.
However, the results for the mean-square charge radius at those pion mass range were significantly smaller than the experimental value, 
and in most cases these studies invoked the NNLO of ChPT to verify that the extrapolation toward the physical pion mass yields the charge radius consistant 
with experiment. Therefore a calculation closer to the physical point is needed to clarify the situation.

Recently PACS-CS collaboration generated gauge ensembles of 2+1 dynamical flavors toward the physical point
using the O(a) improved Wilson-clover quarks and Iwasaki gauge action on a $32^3\times64$ lattice volume with a lattice spacing of 0.0907(13) fm \cite{pacscs}.
Hopping parameters were chosen such that the pion mass covered the range from $m_\pi \approx 702$~MeV down to $m_\pi \approx 156$~MeV.
It was shown that the spectrum quantities can be measured close to the physical point.
This work encourages us to proceed to a project of calculation of the pion form factor on the PACS-CS gauge ensembles. 
This paper is a progress report of our work.

\section{Twisted boundary condition}

The minimum non-zero quark momentum for the periodic boundary condition on an $32^3\times 64$ lattice with a 2GeV inverse lattice spacing is about 0.4GeV.  
To probe the region of smaller momentum transfer one can apply the technique of twisted boundary 
condition \cite{bedaque2004,divitiis2004,sachrajda2005,bedaque2005}.
If one imposes the boundary condition given by 
\begin{equation}
\psi(x+Le_j)=e^{2\pi i\theta_j}\psi (x), \qquad j=1,2,3
\end{equation}
on quark fields 
where $L$ denotes the spatial lattice size and $e_j$ the unit vector in the spatial j-th direction, the spatial momentum is quantized according to 
\begin{equation}
p_j = \frac{2\pi n_j}{L} + \frac{2\pi \theta_j}{L}, \qquad j=1,2,3.
\end{equation}
In this way one can explore arbitrarily small momentum on the lattice by adjusting the value of twist $\theta_j$. 
In practice we transfer the twist from the quark sector to the gluon sector by an $U(1)$ transformation given by
\begin{equation}
\psi(x) \longrightarrow U(\theta,x) \psi(x) = e^{2\pi i \sum^{3}_{j=1}{\theta_j x_j/L}} \psi(x).
\end{equation}
Quark propagators are solved with the periodic boundary condition but on the gluon fields twisted by the U(1) transoformation above. 

\begin{figure}[t]
\centering
\hspace{-55pt}
\subfigure[Pion effective energies.]{\includegraphics[scale=.43]{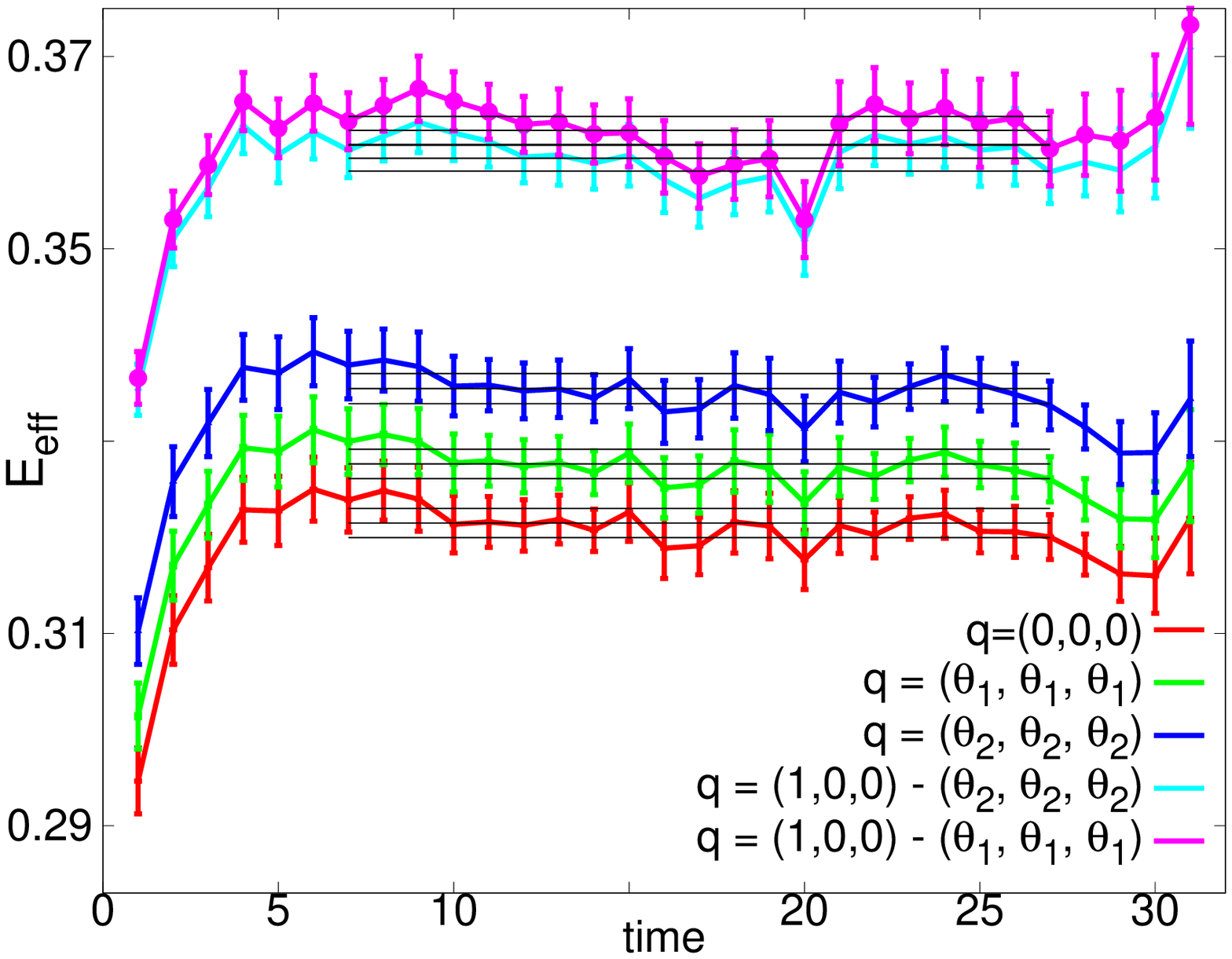}}
\hspace{-25pt}
\subfigure[Continuum dispersion relation.]{\includegraphics[scale=.43]{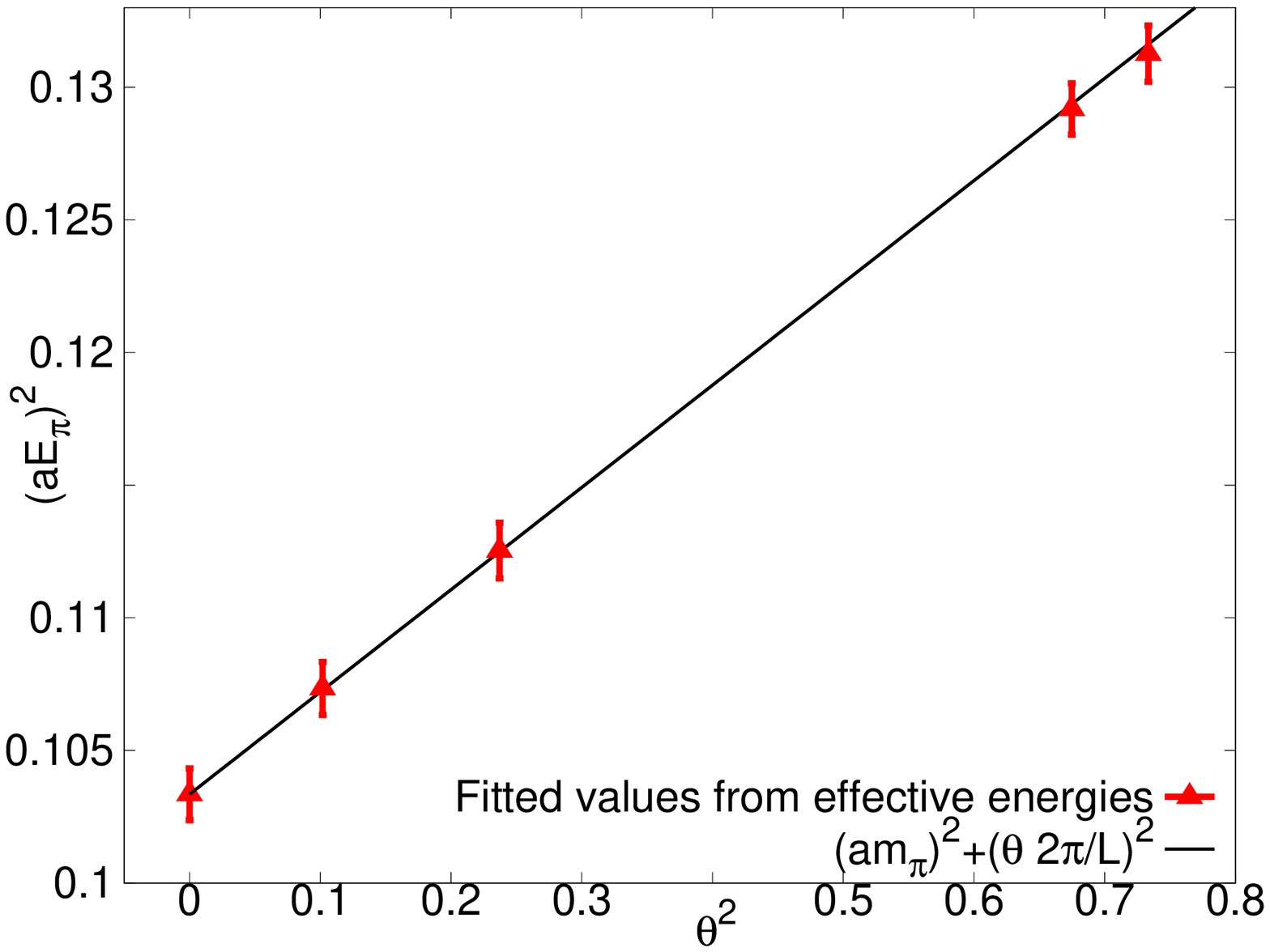}}
\hspace{-65pt}
\caption{Check of validity of the twisted boundary condition at $\kappa_{s}=0.1364, \kappa_{ud}=0.13700$ where $m_\pi \approx 702$~MeV
with 40 configurations measured.}
\label{fig:fig1}
\end{figure}

To check that the term $\frac{2\pi \theta_j}{L}$ acts as a true physical momentum, we carried out a test on the relativity dispersion relation of the pion.
Of the two valence quarks inside the pion, we twisted one quark with a twist angle $\vec{\theta}=(\theta, \theta, \theta)$ and left the other untwisted.
In Fig.\ref{fig:fig1}(a) we plot the effective energy for several values of the twist angle and integer momenta 
at the hopping parameters $\kappa_{s}=0.1364$, $\kappa_{ud}=0.13700$ where $m_\pi \approx 702$~MeV. The propagator is fitted 
over $t=7-27$ to extract the energy $E(\vec p)$.  The error is estimated by the jackknife method with the bin size of 100 trajectories.  
The results are plotted in Fig.\ref{fig:fig1}(b), together with the expected behavior,
\begin{equation}
E\left(\vec p \right)^2 = \left( am_\pi \right)^2 + \left( \frac{2\pi}{L}\vec n + \frac{2\pi}{L}\vec\theta \right)^2, 
\end{equation}
which demonstrates clearly that the term $\frac{2\pi \theta_j}{L}$ acts as a true physical momentum.

\section{Electromagnetic form factor of the pion}
The electromagnetic form factor of the pion is defined by
\begin{equation}
\left< \pi^+(\vec{p'})|V_4|\pi^+(\vec{p})\right> = (p_4 + p'_4)G_\pi(Q^2),
\end{equation}
where $p=(\vec{p},p_4)$ is the 4-momentum of the incoming pion, $p'=(\vec{p'},p'_4)$ that of the outgoing pion, $Q^2=-(p'-p)^2$ the 4-momentum transfer,
and $V_4 = \frac{2}{3}\bar{u}\gamma_4 u-\frac{1}{2}\bar{d}\gamma_4 d$ the electromagnetic current.
Since we employ the local (non-conserved) current for $V_4$, the bare matrix element needs to be renormalized by requiring $G_\pi(0)=1$.

One can extract the form factor from a suitable ratio of two- and three-point functions. In this work, we use the following ratio which 
has the advantage of simultaneously reducing flucutuations and renormalizing the current, 
\begin{equation}
R(\tau)=\frac{C^{3pt}(\vec{p'},t_f;\vec{p},0;\tau)}{C^{3pt}(\vec{p'},t_f;\vec{p'},0;\tau)} \frac{C^{2pt}(\vec{p'},\tau)}{C^{2pt}(\vec{p},\tau)}
\times \frac{2E_\pi(\vec{p'})}{E_\pi(\vec{p})+E_\pi(\vec{p'})}
\xRightarrow{\mbox{large $\tau$ and $t_f$}}\frac{G^{bare}_\pi(Q^2)}{G^{bare}_\pi(0)}=G_\pi(Q^2),
\label{eq:ratio}
\end{equation}
where $C^{2pt}(\vec{p},\tau)=\left< \pi^+(\vec{p},\tau)\pi^+(\vec{p},0)\right>$ and 
$C^{3pt}(\vec{p'},t_f;\vec{p},0;\tau)=\left< \pi^+(\vec{p'},t_f)|V_4(\tau)|\pi^+(\vec{p},0)\right>$ 
are the two- and three-point functions, respectively.  

\begin{figure}[t]
\centering
\hspace{-55pt}
\subfigure[Ratio $R(\tau)$ at various momenta transfer.]{\includegraphics[scale=.43]{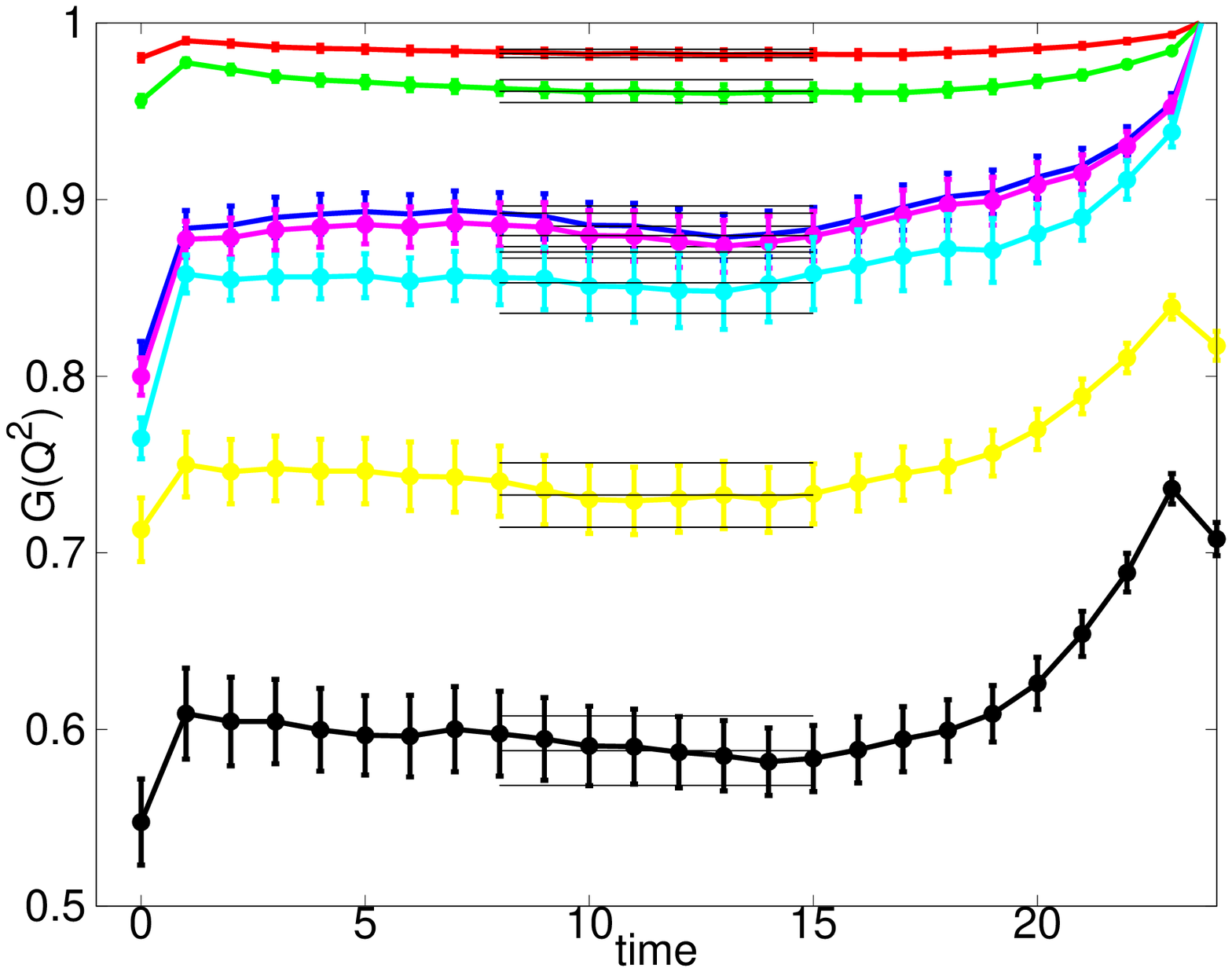}}
\hspace{-25pt}
\subfigure[$Q^2$ dependence of $G_\pi(Q^)$]{\includegraphics[scale=.43]{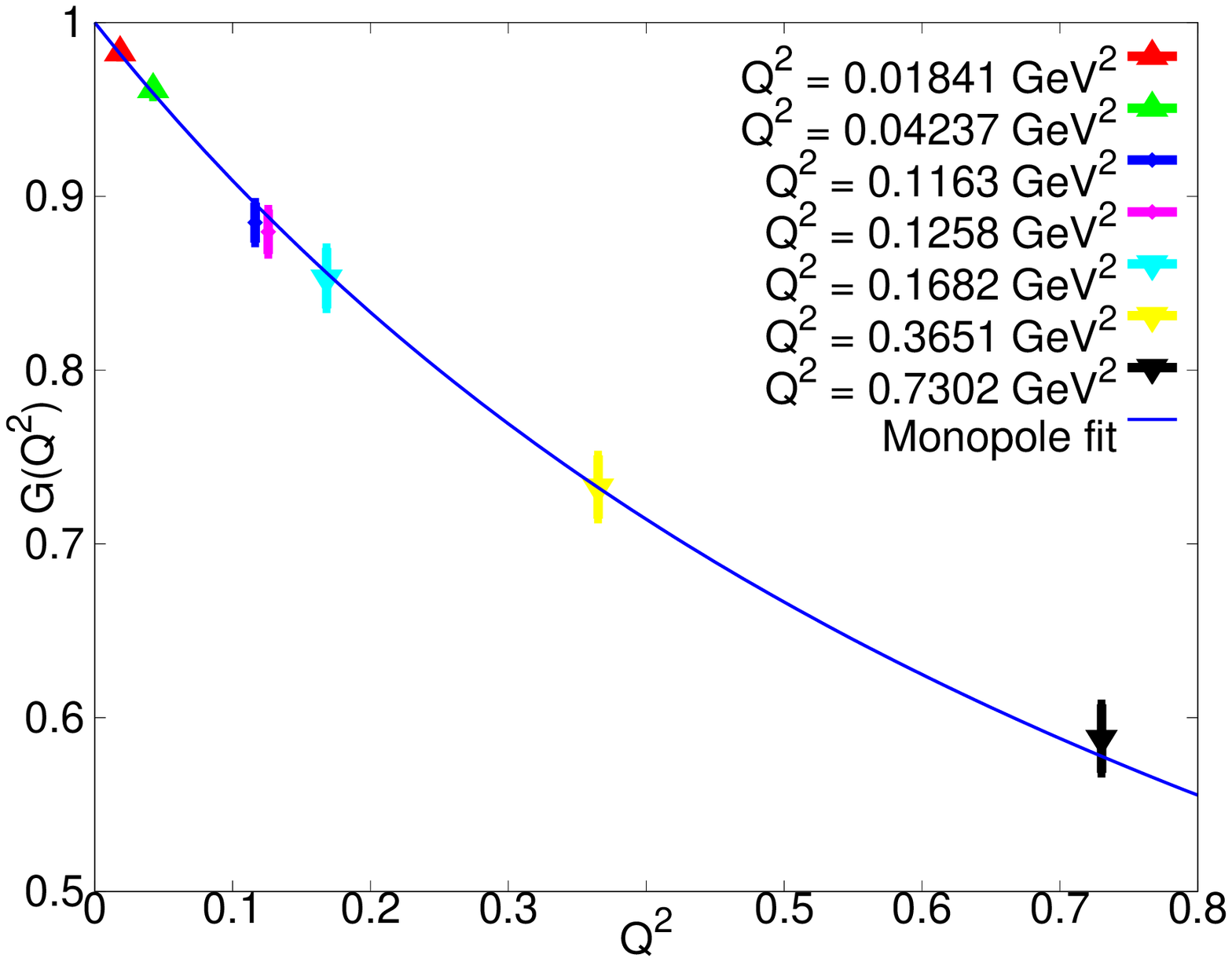}}
\hspace{-65pt}
\caption{Pion form factor at $m_\pi \approx 702$~MeV. (a) the ratio $R(\tau)$ as a function of the time for the current. 
The sink is fixed at $t_f=24$, and the fitting range  of $G_\pi(Q^2)$ is from 8 to 15. (b) Fitted values of $G_\pi(Q^2)$ and a line from the monopole fit.}
\label{fig:fig2}
\end{figure}
\begin{figure}[t]
\centering
\hspace{-55pt}
\subfigure[Ratio $R(\tau)$ at two momenta transfer.]{\includegraphics[scale=.43]{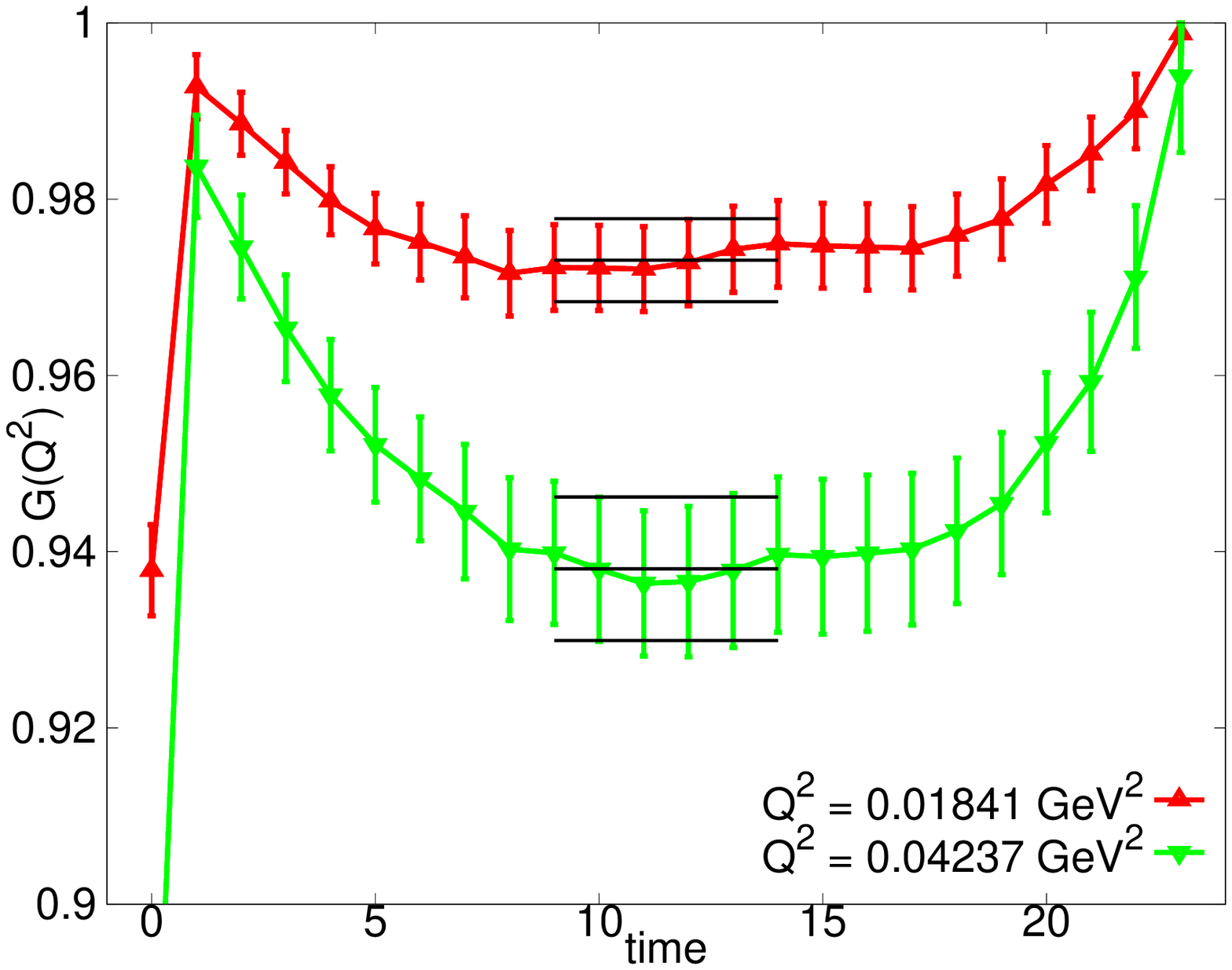}}
\hspace{-25pt}
\subfigure[$Q^2$ dependence of $G_\pi(Q^)$]{\includegraphics[scale=.43]{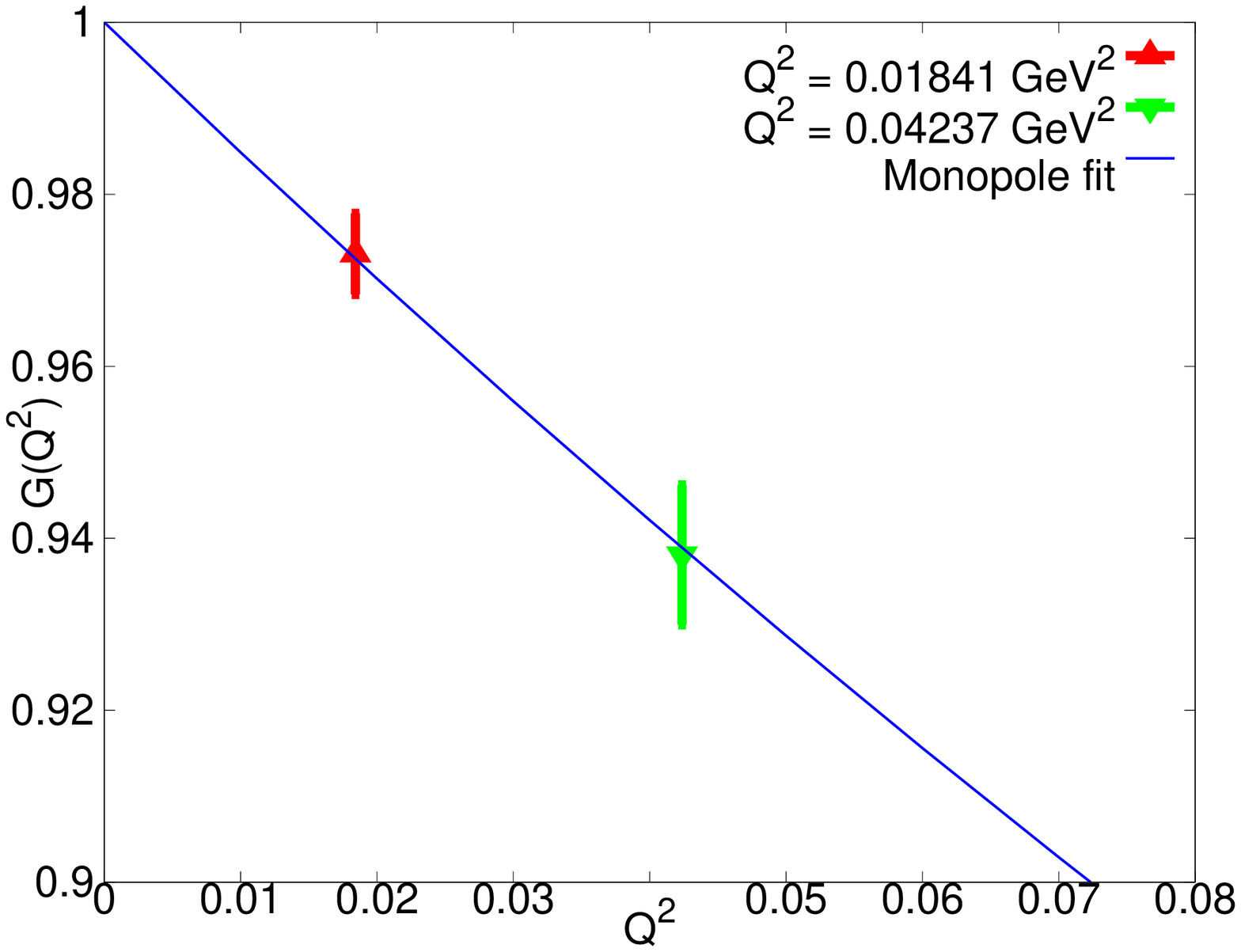}}
\hspace{-65pt}
\caption{The pion form factor at $m_\pi \approx 296$~MeV. The sink time is fixed at $t_f=24$. The fitting range for $G_\pi(Q^2)$ is from 9 to 14.}
\label{fig:fig3}
\end{figure}

\begin{table}[bht]
\centering
\begin{tabular}{c c c c c}
\hline
$\kappa_{s}$ & $\kappa_{s}$ & $M_\pi$ (MeV) & \#conf measured & $\theta$ \\
\hline
0.1364 & 0.13700 & 702 & 40 & 0.18423, 0.28112\\
       & 0.13727 & 570 & 40 & 0.18467, 0.28265\\
       & 0.13754 & 411 & 47 & 0.18585, 0.28672\\
       & 0.13770 & 296 & 160 & 0.18814, 0.29450\\
\hline
\end{tabular}
\caption{Simulation parameters. We adjust $\theta$ values to have the same two smallest $Q^2$ values at every $m_\pi$}
\label{tab:table}
\end{table}

Measurements were made for five values of the degenerate up-down hopping parameter $\kappa_{ud} = \{0.13700$, $0.13727$, $0.13754$, $0.13770$,$0.13781\}$.
The hopping parameter of strange quark is fixed at $\kappa_s = 0.1364$.
The pion mass varies from $M_\pi \approx 702$~MeV down to $M_\pi \approx 156$~MeV.
We fix the final pion at zero momentum $\vec{p'}=\vec{0}$ and vary that of the initial pion $\vec{p}$. 
The  twist technique is applied to the quark running from source to the current. 
Two values were chosen for the twist angle $\vec{\theta}=\left(\theta, \theta, \theta\right)$ such that 
the smallest 4-momentum transfer of the current took the value $Q^2({\rm GeV})=0.01841$ or $0.04237$.  Adding integer momenta, we then collected data 
for $Q^2$ in the range $0.1 {\rm ~GeV} \lesssim Q^2 \lesssim 0.7{\rm ~GeV}$.
We employ an exponentially smeared source and local sink. The three-point function was calculated by the conventional source method \cite{martinelli,draper}.
The simulation parameters are listed in Table \ref{tab:table}.  

In Fig.\ref{fig:fig2} we depict data for the pion form factor at pion mass $m_\pi \approx 702$~MeV.
To extract the form factor, we fit the plateau of the ratio $R(\tau)$ by a constant.
The fitting range should be chosen around the symmetry point between the source and the sink, with additional 
considerations on the time interval required for the pion state to become dominant. 
In our case the sink is fixed at $t_f=24$, thus the fitting should be symmetric around $t=12$ if the sink and the source are of the same kind.
However, since we use a smeared source and the point sink, we shift the fitting range one time unit to the source, and 
choose $\tau=8$ to 15.

Figure \ref{fig:fig3} shows similar data at the pion mass $m_\pi \approx 296$~MeV.  
At current statistics, we have reasonable signal for only two smallest momenta transfer $Q^2=\{0.01841, 0.04237\}$.

\section{Mononople ansatz and mean-square charge radius of the pion}

The experimental pion form factor follows the monopole form suggested by the vector dominance model
\begin{equation}
G_\pi(Q^2) = \frac{1}{1+{Q^2/M^2_{mono}}}.
\end{equation}
Our data confirms this trend as illustrated in Fig.\ref{fig:fig2}.
we then fit our data to the monopole form from which we extract the mean-square charge radius of the pion
 \begin{equation}
 \left<r^2\right> \equiv 6\frac{dG_\pi(Q^2)}{dQ^2}|_{Q^2=0}.
 \end{equation}

\begin{figure}
\begin{center}
\includegraphics[scale=.6]{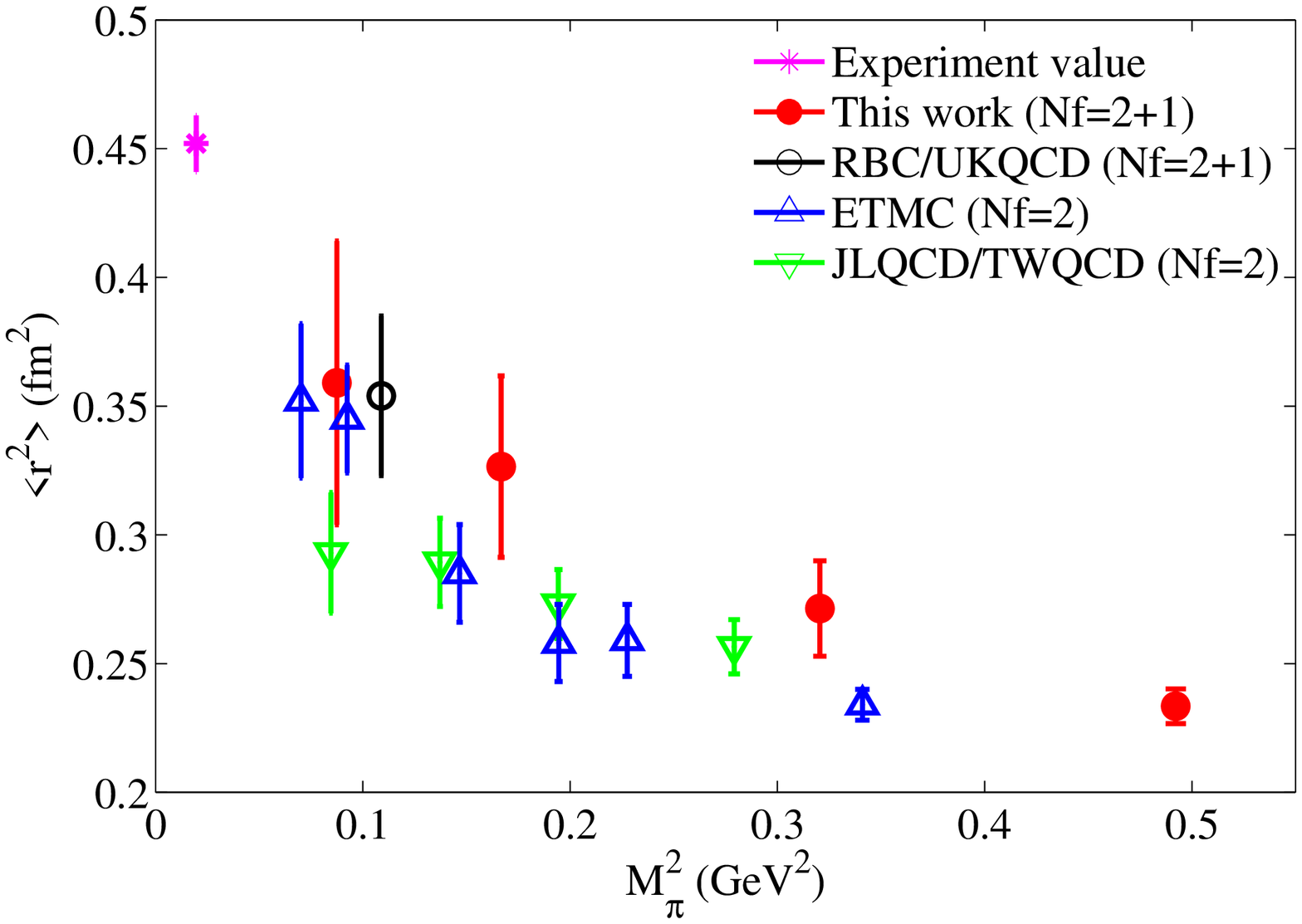}
\end{center}
\caption{$\left< r^2 \right>$ dependence on $M^2_\pi$}
\label{fig:radius}
\end{figure}
In Fig.\ref{fig:radius} we plot our initial results (filled circles) as a function of pion mass squared, 
together with the experiment value (burst) and those from latest researches: the only work for 2+1 dynamical flavor QCD is from RBC/UKQCD Collaboration 
at a single pion mass (open cirlce) using domain-wall quark acton, and others are for 2 dynamical flavors from ETMC Collaboration (twisted mass QCD; triangles) and from 
JLQCD (overlap action; inverted triangles).  

\section{Conclusions}

We have presented our initial results of a lattice calculation of the pion electromagnetic form factor
in 2+1 dynamical flavor QCD with the O(a) improved Wilson-clover quarks and Iwasaki gauge action.
Our current data of $\left< r^2 \right>$ at small pion masses agree with the recent data from other groups, and 
$\left< r^2 \right>$ show an increasing behavior toward the physical value as $M_\pi$ decreases.

At present our data only goes down to $M_\pi\approx 296$~MeV whereas the PACS-CS gluon ensemble reaches $M_\pi\approx 156$~MeV.  Reducing the pion mass to 156~MeV, 
we have observed an increasingly larger fluctuation in the two- and three-point functions, and this trend worsens for larger momenta.  
For having better statistics, we are currently exploring various techniques. One of the possibilities is changing the set up to the channel where $|\vec{p'}|=|\vec{p}|$.
In this channel, the two-point function in the  ratio (\ref{eq:ratio}) drops out, leaving just the ratio of three-point functions.
This ratio of the three-point function with non-zero momentum transfer over that with zero momentum transfer seems to
give signals with smaller statistical error compared with the ratio we used in the present calculation.
We are also considering to use the random noise technique to reduce computer time for calculation of the two- and three-point functions.

\section*{Acknowledgements}

The calculations reported here have been carried out on the PACS-CS and T2K-Tsukuba computers under the ``Interdisciplinary Computational Science Program'' 
of Center for Computational Sciences of University of Tsukuba.  This work is supported in part by Grants-in-Aid for Scientific Research from the Ministry of 
Education, Cultute, Sports, Science and Technology (Nos. 18104005).

\end{document}